\documentclass[12pt]{article}
\usepackage[cp1251]{inputenc}
\usepackage[russian, english]{babel}
\usepackage{amssymb}

\newcommand{\bm}[1]{\mbox{\boldmath{$#1$}}}

\begin{document}

\thispagestyle{empty}

\begin{center}
{\bf ON THE LAGRANGE STABILITY OF MOTION AND  FINAL EVOLUTIONS IN
THE THREE-BODY PROBLEM}
\end{center}

\begin{center}
{\large S.P. Sosnitskii}\\
{\small {\it Institute of Mathematics of Ukrainian National
Academy of Sciences,\\
Tereshchenkivs'ka  str 3, 01601, MSP, Kyiv--4, Ukraine,\\
E-mail address: sosn@imath.kiev.ua}}\\
\vspace{0.3cm}

\end{center}

\noindent {\bf Abstract.} {\small For the three-body problem, we
consider the Lagrange stability. To analyze the stability, along
with integrals of energy and angular momentum, we use relations by
the author from \cite{Sosn2}, which band together separately
squared mutual distances between bodies (mass points) and squared
mutual distances from bodies to the barycenter of the system. In
this case, we prove the Lagrange stability theorem, which allows
us to define more exactly the character of  hyperbolic-elliptic
and parabolic-elliptic final evolutions.}



\section{Introduction}

It is known \cite {Whitt,Dub,Pars} that the three-body problem
(for mass points) is considered for the system of three bodies
with masses $m_1, m_2, m_3,$ respectively, that are in the
movement in the three-dimensional Euclidean space under the mutual
gravitational attraction. We have to determine their coordinates
and velocities at any time $t$ on the base of initial data. In
this form, despite of significant progress based on the
achievements of Kolmogorov-Arnold-Moser theory \cite {Arnold1},
the problem remains unsolved until now, and therefore a
qualitative study of motion in this system is still important. In
particular, it is still important to obtain an answer for the
following question: What are conditions under which three bodies
remans inside a bounded domain of the Euclidean space. Later, we
will suggest sufficient conditions for the boundedness of the
motion.

Before we start to investigate the motion of the mass points, we write down the formula for the related Lagrangian:
$$
L = T + U = \frac {1}{2} \sum_i ^3  m_i {\bf \dot r}_i^2 + G \left(\frac {m_1 m_2}{|{\bf r}_{12}|}+ \frac {m_1 m_3}{|{\bf
r}_{13}|}+ \frac {m_2 m_3}{|{\bf r}_{23}|}\right). \eqno(1.1)
$$
Here, ${\bf r}_i$ are radius vectors of points in the inertial reference system with the origin at the center of masses $m_i$, ${\bf r}_{ij} = {\bf r}_j -
{\bf r}_i, \  (i,j = 1,2,3)$, \ $G>0$ is the gravitation constant.
The motion equations for the Lagrangian (1.1) take the following form
$$
\ddot{\bf r}_{1}= G\biggl( m_2 \frac {{\bf r}_2 - {\bf r}_{1}} {|{\bf r}_{12}|^3} + m_3 \frac {{\bf r}_3 -{\bf r}_{1}} {|{\bf r}_{13}|^3} \biggl),
$$
$$
\ddot{\bf r}_{2}= G\biggl( -m_1 \frac {{\bf r}_2 - {\bf r}_{1}} {|{\bf r}_{12}|^3} + m_3 \frac {{\bf r}_3 -{\bf r}_{2}} {|{\bf r}_{23}|^3} \biggl), \eqno(1.2)
$$
$$
\ddot{\bf r}_{3}= G\biggl( {-m_1} \frac {{\bf r}_3 - {\bf r}_{1}} {|{\bf r}_{13}|^3} - m_2 \frac {{\bf r}_3 -{\bf r}_{2}} {|{\bf r}_{23}|^3} \biggl).
$$
Passing over to dimensionless time variable
$t{\sqrt{GM}}/{r_0^{3/2}}=\tau$ in (1.2), where  $M=m_1+m_2+m_3$
and $r_0$ is a parameter with the dimension of the length unit, we
obtain the following equations \cite {Sosn3}
$$
{\bm \rho}_{1}''=  \mu_2 \frac {{\bm \rho}_2 - {\bm \rho}_{1}} {|{\bm \rho}_{12}|^3} + \mu_3 \frac {{\bm \rho}_3 -{\bm \rho}_{1}} {|{\bm \rho}_{13}|^3},
$$
$$
{\bm \rho}_{2}''= -\mu_1 \frac {{\bm \rho}_2 - {\bm \rho}_{1}} {|{\bm \rho}_{12}|^3} + \mu_3 \frac {{\bm \rho}_3 -{\bm \rho}_{2}} {|{\bm \rho}_{23}|^3}, \eqno(1.3)
$$
$$
{\bm \rho}_{3}''=  {-\mu_1} \frac {{\bm \rho}_3 - {\bm \rho}_{1}} {|{\bm \rho}_{13}|^3} - \mu_2 \frac {{\bm \rho}_3 -{\bm \rho}_{2}} {|{\bm \rho}_{23}|^3}.
$$
Here, the prime sign denotes the differentiation with respect to $\tau$, $\mu_i=m_i/M$, ${\bm\rho_i}={{\bf r_i}}/{r_0}$ are relative radius vectors.

In what follows, along with equations (1.3), we will use the
following equations for distances that were obtained in \cite
{Sosn3}:
$$
{\rho_{12}^2}''=2v_{12}^2 - 2\frac{\mu_1+\mu_2}{\rho_{12}} + \frac{\mu_3}{\rho_{13}}
\left(\frac{\rho_{23}^2-\rho_{12}^2}{\rho_{13}^2}-1\right)+ \frac{\mu_3}{\rho_{23}}
\left(\frac{\rho_{13}^2-\rho_{12}^2}{\rho_{23}^2}-1\right),
$$
$$
{\rho_{13}^2}''=2v_{13}^2 -2\frac{\mu_1+\mu_3}{\rho_{13}}+\frac{\mu_2}{\rho_{12}}
\left(\frac{\rho_{23}^2-\rho_{13}^2}{\rho_{12}^2}-1\right)+ \frac{\mu_2}{\rho_{23}}
\left(\frac{\rho_{12}^2-\rho_{13}^2}{\rho_{23}^2}-1\right),
$$
$$
{\rho_{23}^2}''=2v_{23}^2 -2\frac{\mu_2+\mu_3}{\rho_{23}}+\frac{\mu_1}{\rho_{12}}
\left(\frac{\rho_{13}^2-\rho_{23}^2}{\rho_{12}^2}-1\right)+ \frac{\mu_1}{\rho_{13}}
\left(\frac{\rho_{12}^2-\rho_{23}^2}{\rho_{13}^2}-1\right),
$$
$$
{v_{12}^2}'+\frac{{\rho_{12}^2}'} {\rho_{12}^3}=\mu_3\left[{\rho_{12}^2}' \left(\frac{1}{\rho_{12}^3}-\frac{1}{\rho_{13}^3}\right)+\right.
$$
$$
\left. +{\rho_{23}^2}' \left(\frac{1}{\rho_{13}^3}-\frac{1}{\rho_{23}^3}\right)+ 2{\bm \rho}_{23}{\bm \rho}_{13}'\left(\frac{1}{\rho_{23}^3}-
\frac{1}{\rho_{13}^3}\right)\right], \eqno(1.4)
$$
$$
{v_{13}^2}'+\frac{{\rho_{13}^2}'}
{\rho_{13}^3}=-\mu_2\left[{\rho_{13}^2}'
\left(\frac{1}{\rho_{12}^3}-\frac{1}{\rho_{13}^3}\right)+
 2{\bm \rho}_{23}{\bm \rho}_{13}'\left(\frac{1}{\rho_{23}^3}-
\frac{1}{\rho_{12}^3}\right)\right],
$$
$$
{v_{23}^2}'+\frac{{\rho_{23}^2}'}
{\rho_{23}^3}=\mu_1\left[{\rho_{23}^2}'
\left(\frac{1}{\rho_{23}^3}-\frac{1}{\rho_{12}^3}\right)+\right.
$$
$$
\left. +2({\bm \rho}_{13}{\bm \rho}_{23})'
\left(\frac{1}{\rho_{12}^3}-\frac{1}{\rho_{13}^3}\right)- 2{\bm
\rho}_{23}{\bm \rho}_{13}'\left(\frac{1}{\rho_{12}^3}-
\frac{1}{\rho_{13}^3}\right)\right],
$$
$$
({\bm \rho}_{23}{\bm \rho}_{13}')'=\frac{1}{2}(-v_{12}^2+v_{13}^2
+ v_{23}^2)-
\frac{\mu_1+\mu_3}{2\rho_{13}}\left(1+\frac{\rho_{23}^2-\rho_{12}^2}
{\rho_{13}^2}\right)+
$$
$$
+\frac{\mu_2}{2\rho_{12}}\left(1+\frac{\rho_{23}^2-\rho_{13}^2}
{\rho_{12}^2}\right)-\frac{\mu_2}{\rho_{23}},
$$
where $ \rho_{ij}=|{\bm \rho}_{ij}|, \  v_{ij}=|{\bm \rho}_{ij}'|$.

The system of ten equations (1.4) is an integral manifold (i.e., a subset) of system (1.3) and it is useful in the study of orbital stability of motions.

In what follows, we will also use the integral of energy
$$
\frac {1}{2} \sum_i^3  \mu_i {\bm \rho}_{i}'^2 - \sum_{i<j}\frac {\mu_i \mu_j}{|{\bm \rho}_{ij}|} = h =\mbox{const} \eqno(1.5)
$$
and the vector integral of angular momentum
$$
\sum_{i}^{3} \mu_i ({\bm \rho}_i\times {\bm \rho}_i') = {\bf C}.
\eqno(1.6)
$$
Next, we will  always assume that ${\bf C}\neq {\bf 0}$.

Since, additionally, there are integrals of motion for the center
of mass for this system, without loss of generality in what
follows we can assume in accordance with the choice of coordinate
system that
$$
\sum_{i}^{3} \mu_i {\bm \rho}_i' = {\bf 0},\ \ \sum_{i}^{3} \mu_i {\bm \rho}_i = {\bf 0},                    \eqno(1.7)
$$
and, as a consequence \cite {Dub, R, Arnold2},
$$
\sum_{i}^{3} \mu_i {\bm \rho}_i^2 = \sum_{i<j} \mu_i \mu_j |{\bm \rho}_{ij}|^2.  \eqno(1.8)
$$

Finally, we will also use obtained  in  \cite {Sosn2}, as a
consequence of (1.7), the following equations:
$$
{\rho}^2_{1} =\mu_2 (\mu_2+\mu_3) {\rho}^2_{12} +
\mu_3(\mu_2+\mu_3) {\rho}^2_{13} - \mu_2\mu_3 {\rho}^2_{23},
$$
$$
{\rho}^2_{2} = \mu_1 (\mu_1+\mu_3) {\rho}^2_{12} - \mu_1\mu_3 {
\rho}^2_{13} + \mu_3(\mu_1+\mu_3) {\rho}^2_{23}, \eqno(1.9)
$$
$$
{\rho}^2_{3} = - \mu_1\mu_2 {\rho}^2_{12} + \mu_1 (\mu_1+\mu_2) {
\rho}^2_{13} + \mu_2(\mu_1+\mu_2) {\rho}^2_{23}.
$$
By reversing equations (1.9), we have
$$
{\rho}^2_{12}=\frac{\mu_1(\mu_1+\mu_2){\rho}_1^2+\mu_2(\mu_1+\mu_2){\rho}_2^2
-\mu_3^2{\rho}_3^2} {\mu_1\mu_2},
$$
$$
{\rho}^2_{13}=\frac{\mu_1(\mu_1+\mu_3){\rho}_1^2-\mu_2^2{\rho}_2^2+
\mu_3(\mu_1+\mu_3){\rho}_3^2} {\mu_1\mu_3},  \eqno(1.10)
$$
$$
{\rho}^2_{23}=\frac{-\mu_1^2 {\rho}_1^2
+\mu_2(\mu_2+\mu_3){\rho}_2^2+\mu_3(\mu_2+\mu_3){\rho}_3^2}
{\mu_2\mu_3}.
$$
Here $\rho_{i}=|{\bm \rho}_{i}|, \rho_{ij}=|{\bm \rho}_{ij}||$.
Similar equations connect ${{\bm \rho}_i'}^2$  and  ${{\bm
\rho}_{ij}'}^2$ \cite {Sosn2}.

\section{Main Definitions and Assumptions}

{\bf Definition  1.} We say that the motion ${\bm \rho}(\tau) =
({\bm \rho}_{1},{\bm \rho}_{2}, {\bm \rho}_{3})^T$ of system (1.3)
 is {\it Lagrange stable} if the following condition is satisfied:
$$
c_1 \le |{\bm \rho}_{ij}(\tau)| \le c_2  \ \ \ \forall \tau \in R = ]- \infty, \infty [,  \ \ \ \forall  i < j, \eqno(2.1)
$$
where $c_1, c_2$ are positive constants.

{\bf Definition  2.} We say that the motion ${\bm \rho}(\tau) =
({\bm \rho}_{1},{\bm \rho}_{2}, {\bm \rho}_{3})^T$ of system (1.3) is {\it distal}
if the following inequality is satisfied:
$$
|{\bm \rho}_{ij}(\tau)| \geq c_3  \ \ \ \forall \tau \in R ,
 \ \ \ \forall  i < j , \ \ \ 0<c_3=\mbox{const}. \eqno(2.2)
$$

As it was mentioned above, equations (1.3) contain relative radius vectors ${\bm\rho_i}={\bf r}_i/{r_0}$ where  $r_0$ is a parameter that has the dimension of the length unit.
Therefore, without loss of generality in what follows, it is convenient for us to put $r_0$ at a value, for which we have $c_1 = c_3 = 1$ in inequalities (2.1) and (2.2).

{\bf Definition  3.} In accordance with \cite{Gol}, we say that a fixed pair of points $(\mu_i,\mu_j),$ $i<j,$ of system (1.3) is {\it Hill stable} if the following inequality is satisfied:
$$
|{\bm \rho}_{ij}(\tau)| < c_4 \ \ \ \forall \tau \in R, \
0<c_4=\mbox{const}. \eqno(2.3)
$$

{\bf Definition  4.} In accordance with \cite{Gol}, we say that a
fixed pair of mass points $(\mu_i,\mu_j),$ $i<j$, of system (1.3)
is {\it Hill absolutely stable} if the following inequality is
satisfied:
$$
\frac{|\bf R(\tau)|}{|{\bm \rho}_{ij}(\tau)|}>
\frac{\mbox{max}(\mu_i,\mu_j)}{\mu_i+\mu_j} \ \ \ \forall \tau \in
R, \eqno(2.4)
$$
where  $|\bf R(\tau)|$  denotes distance from third mass  point to
the center of mass of fixed pair of points $(\mu_i,\mu_j)$.

As it is proved in  \cite{Gol}, if a fixed pair of mass points
$(\mu_i,\mu_j),$ $i<j$, of system (1.3) is  Hill absolutely
stable, then it is Hill stable and collisions are possible only
for mass points, which form this fixed pair.

Key points for forming of initial conditions, under which we have
the Hill stability of a pair of mass points, are integrals of
energy and angular momentum \cite {Gol, March, Luk}.

{\bf Lemma 1.} {\it If one of the pairs of mass points in the
three-body problem is Hill stable, then there exists a closed ball
$\bar{B_{r}}$ in the appropriate configuration space $R^{9}$ such
that none of the vectors ${\bm\rho}_{i}$ in
$R^{9}\backslash\bar{B_{r}}$ can be a zero vector}.

{\bf Proof.} The lemma is obvious when it comes to the triple
collision. Therefore, in what follows, we restrict ourselves to
the case where only one of the vectors ${\bm\rho}_{i}$ is a zero
vector.

As it is known (see e.g. \cite{Sosn1}), the following relations
are valid:
$$
{\bm \rho}_{1}=-\mu_{2}{\bm \rho}_{12}-\mu_{3}{\bm \rho}_{13},
$$
$$
{\bm \rho}_{2}=\mu_{1}{\bm
\rho}_{12}-\mu_{3}{\bm\rho}_{23},\eqno(2.5)
$$
$$
{\bm \rho}_{3}=\mu_{1}{\bm \rho}_{13}+\mu_{2}{\bm \rho}_{23}.
$$
Suppose that ${\bm \rho}_{1}={\bf 0}$. Then due to the first
relation of system (2.5) we have
$$
\mu_{2}{\bm \rho}_{12}+\mu_{3}{\bm \rho}_{13}={\bf 0}.\eqno(2.6)
$$
Supplementing equality (2.6) with the identity
$$
{\bm \rho}_{12}- {\bm \rho}_{13}=-{\bm \rho}_{23},\eqno(2.7)
$$
we obtain
$$
{\bm \rho}_{12}=-\frac{\mu_{3}}{\mu_{2}+\mu_{3}}{\bm \rho}_{23}, \
\ {\bm \rho}_{13}=\frac{\mu_{2}}{\mu_{2}+\mu_{3}}{\bm
\rho}_{23},\eqno(2.8)
$$
and these relations show that if at least one of the distances
${\bm\rho}_{ij}$ is bounded, then all three distances are bounded.

If  we have either the equality ${\bm \rho}_{2}={\bf 0}$ or the
equality ${\bm\rho}_{3}={\bf 0}$ instead of ${\bm \rho}_{1}={\bf
0}$, we argue similarly.

In what follows, without loss of generality, we assume that the
Hill stable pair is the pair $(\mu_{1},\mu_{2})$. Then, by using
equalities (1.9), in dependence of which one of the vectors ${\bm
\rho}_{i}$ is a zero vector, we obtain three different expressions
for the radius of the ball that is referred to the center of mass
of three particles:
$$
r_{i}^{2}=\sum_j ^3{{\bm \rho}_{j}^{2}}=
{\bm\rho}_{12}^{2}f_{i}({\bm \mu}), \ (i=1,2,3), \ \
{\bm\rho}_{j}={\bf 0} \ \ \forall j =i, \ \
{\bm\mu}=(\mu_{1},\mu_{2},\mu_{3})^{T}. \eqno(2.9)
$$
Equalities (2.9) allow us to conclude that if one of the vectors
${\bm \rho}_{i}$ is zero vector, then motions can be embedded into
a closed ball $\bar{B_{r}}$ with the radius defined by relations
$$
|{\bm \rho}_{12}|^{*}=\mbox{sup}(|{\bm \rho}_{12}|),\ \
f^{*}=\mbox{max}(f_{1},f_{2},f_{3}). \eqno(2.10)
$$
The lemma 1 is proved. $\Box$

{\bf Corollary 1}. {\it The scheme of the proof of Lemma 1 implies
that the radius $r$ of the sphere  $\bar{B_{r}}$ can always be
chosen not only in such a way that each of the variables
${\rho}_{i}=|{\bm \rho}_{i}|$ is not vanish in
$R^{9}\backslash\bar{B_{r}}$, but also to exceed some positive
constant}.

{\bf Corollary 2}. {\it If the motion in the three-body problem is
outgoing, then surely there is a time $\tau^{*}$ such that the
segment of the orbit (the projection of the phase trajectory in
the configuration space) falls into $R^{9}\backslash\bar{B_{r}}$
for $\tau >\tau^{*}$}.

{\bf Lemma 2}. {\it Let ${\bm \rho}(\tau) =({\bm \rho}_{1},{\bm
\rho}_{2}, {\bm \rho}_{3})^T$ be a  Lagrange unstable motion of
system (1.3), for which   the pair of bodies $(\mu_1,\mu_2)$  is
Hill  absolutely stable. \noindent

Then, for this  motion, there is a sequence $\{\tau_k\}$
$(k=1,2,3,\dots)$ such that the equalities
$$
\lim_{\tau_k \to
\infty}\left(\frac{{\rho}_2^2(\tau_k)}{{\rho}_1^2(\tau_k)}\right)=1,
\ \ \lim_{\tau_k \to
\infty}\left(\frac{{\rho}_3^2(\tau_k)}{{\rho}_1^2(\tau_k)}\right)=
\frac{(\mu_1+\mu_2)^2}{\mu_3^2} \eqno(2.11)
$$
are valid.}

{\bf Proof.} Since  the motion under consideration is Lagrange
unstable,  there is a sequence $\{\tau_k\}$ $(k=1,2,3,\dots)$ such
that
$$
\lim_{k \to \infty} \tau_k = \infty,  \ \ \ \lim_{k \to \infty}
{\rho}_i(\tau_k)= \infty \ \  \forall i=1,2,3. \eqno(2.12)
$$
Let us divide the first equality of system (1.10) by
${\rho}^2_{1}$. As a result, for the Lagrange unstable motion  we
have
$$
\frac{{\rho}^2_{12}(\tau_k)}{{\rho}_1^2(\tau_k)}=\frac{1}{\mu_1\mu_2}\left[\mu_1(\mu_1+\mu_2)
+\mu_2(\mu_1+\mu_2)\frac{{\rho}_2^2(\tau_k)}{{\rho}_1^2(\tau_k)}
-\mu_3^2\frac{{\rho}_3^2(\tau_k)}{{\rho}_1^2(\tau_k)}\right].
\eqno(2.13)
$$
Tending $k$ to infinity in equality (2.13), we obtain the equality
$$
\mu_2(\mu_1+\mu_2)\left(\frac{{\rho}_2^2}{{\rho}_1^2}\right)_{\infty}
-\mu_3^2\left(\frac{{\rho}_3^2}{{\rho}_1^2}\right)_{\infty}=-\mu_1(\mu_1+\mu_2).
\eqno(2.14)
$$
Further, on the base of last two equalities of system (1.10), we
derive
$$
\left(\frac{\mu_1}{\mu_2}\right)\frac{{\rho}^2_{13}(\tau_k)}{{\rho}^2_{23}(\tau_k)}=
$$
$$
=\frac{\mu_1(\mu_1+\mu_3)-\mu_2^2{{\rho}_2^2(\tau_k)}/{{\rho}_1^2(\tau_k)}+
\mu_3(\mu_1+\mu_3){{\rho}_3^2(\tau_k)}/{{\rho}_1^2(\tau_k)}}{-\mu_1^2
+\mu_2(\mu_2+\mu_3){{\rho}_2^2(\tau_k)}/{{\rho}_1^2(\tau_k)}+
\mu_3(\mu_2+\mu_3){{\rho}_3^2(\tau_k)}/{{\rho}_1^2(\tau_k)}}.\eqno(2.15)
$$

Observing
$$
{\bm \rho}_{13}^2 ={\bm \rho}_{12}^2 + {\bm \rho}_{23}^2 + 2{\bm
\rho}_{12}{\bm \rho}_{23}
$$
and taking into account  (1.8), (2.12), we obtain
$$
\lim_{k \to
\infty}\frac{{\rho}^2_{13}(\tau_k)}{{\rho}^2_{23}(\tau_k)}=
\lim_{k \to
\infty}\left[\frac{{\rho}^2_{12}(\tau_k)}{{\rho}^2_{23}(\tau_k)}+
2\frac{{\rho}_{12}(\tau_k)}
{{\rho}_{23}(\tau_k)}\cos(\widehat{{\bm \rho}_{12}(\tau_k),{\bm
\rho}_{23}}(\tau_k))+1\right]=1. \eqno(2.16)
$$
In the limit, on the base of  (2.15), (2.16), we have
$$
\mu_2[\mu_2^2+\mu_1(\mu_2+\mu_3)]
\left(\frac{{\rho}_2^2}{{\rho}_1^2}\right)_{\infty}+
\mu_3^2(\mu_1-\mu_2)\left(\frac{{\rho}_3^2}{{\rho}_1^2}\right)_{\infty}=
\mu_1[\mu_1^2+\mu_2(\mu_1+\mu_3)]. \eqno(2.17)
$$
By equations (2.14), (2.17) we derive
$$
\left(\frac{{\rho}_2^2}{{\rho}_1^2}\right)_{\infty}=1, \ \
\left(\frac{{\rho}_3^2}{{\rho}_1^2}\right)_{\infty}=
\frac{(\mu_1+\mu_2)^2}{\mu_3^2}.
$$

Lemma 2 is proved.  $\Box$

{\bf Lemma 3}. {\it Let ${\bm \rho}(\tau) =({\bm \rho}_{1},{\bm
\rho}_{2}, {\bm \rho}_{3})^T$ be a distal and Lagrange unstable
motion of system (1.3), for which   the pair of bodies
$(\mu_1,\mu_2)$ is Hill  stable.  \noindent

Then,  there is a sequence $\{\tau_k\}$ $(k=1,2,3,\dots)$ such
that  in the limit case   one of  the equalities
$$
\left\{\frac{{\rho}^2_{13}}{{\rho}^2_{12}} -
\frac{{\rho}^2_{23}}{{\rho}^2_{12}}\right\}_{\infty}=
\frac{\mu_1-\mu_2}{\mu_3}, \eqno(2.18)
$$
$$
\left\{-\frac{{\rho}^2_{13}}{{\rho}^2_{12}}+
\frac{{\rho}^2_{23}}{{\rho}^2_{12}}\right\}_{\infty}=\frac{\mu_1\mu_3^2+
(\mu_2+\mu_3)(\mu_1+\mu_2)^2}{(\mu_1+\mu_2)\mu_3}, \eqno(2.19)
$$
$$
\left\{\frac{{\rho}^2_{13}}{{\rho}^2_{12}} -
\frac{{\rho}^2_{23}}{{\rho}^2_{12}}\right\}_{\infty}=\frac{\mu_2\mu_3^2+
(\mu_1+\mu_3)(\mu_1+\mu_2)^2}{(\mu_1+\mu_2)\mu_3}. \eqno(2.20)
$$
is  valid}.

{\bf Proof.} Since  the motion under consideration is Lagrange
unstable,  there is a sequence $\{\tau_k\}$ $(k=1,2,3,\dots)$ such
that
$$
\lim_{k \to \infty} \tau_k = \infty,
 \ \ \ \lim_{k \to \infty} \ \sum^3_{i<j}{\rho}^2_{ij} (\tau_k)= \infty.
                                                       \eqno(2.21)
$$

We rewrite equalities (1.9) in the following form:
$$
\frac{1}{\mu_2(\mu_2+\mu_3)}u^2 - \frac{\mu_3}{\mu_2}v^2+ \frac{\mu_3}{\mu_2+\mu_3}w^2=1,
$$
$$
\frac{1}{\mu_1(\mu_1+\mu_3)}\left(\frac{{\rho}^2_{2}}{{\rho}^2_{1}}\right)
u^2+ \frac{\mu_3}{\mu_1+\mu_3}v^2- \frac{\mu_3}{\mu_1}w^2=1,
\eqno(2.22)
$$
$$
-\frac{1}{\mu_1\mu_2}\left(\frac{{\rho}^2_{3}}{{\rho}^2_{1}}\right) u^2+ \frac{\mu_1+\mu_2}{\mu_2}v^2+ \frac{\mu_1+\mu_2}{\mu_1}w^2=1,
$$
where
$$
u^2=\frac{{\rho}^2_{1}}{{\rho}^2_{12}}, \
v^2=\frac{{\rho}^2_{13}}{{\rho}^2_{12}}, \
w^2=\frac{{\rho}^2_{23}}{{\rho}^2_{12}}. \eqno(2.23)
$$
As a result, we obtain a system of three equations that are linear
with respect to $u^2, v^2, w^2$ and contain variable coefficients
${\rho}^2_{2}/{\rho}^2_{1}$ and ${\rho}^2_{3}/{\rho}^2_{1}$, and
each one of these equations can be treated as an equation of a
one-sheet hyperboloid. Moreover, if the first equation describes a
stationary hyperboloid, then the second and the third ones
describe movable hyperboloids, if we take into account the fact
that coefficients ${\rho}^2_{2}/{\rho}^2_{1}$ and
${\rho}^2_{3}/{\rho}^2_{1}$ are variable. All these hyperboloids
have distinct imaginary semiaxes.

Let us exclude the variable $u^2$ from equations (2.22). As a
result, we obtain equations
$$
v^2-\frac{\alpha_1}{\beta_1}w^2=\frac{\gamma_1}{\mu_3\beta_1},
$$
$$
v^2+\frac{\alpha_2\mu_2}{\beta_2}w^2=\frac{\mu_2\gamma_2}{\beta_2},
\eqno(2.24)
$$
$$
v^2+\frac{\alpha_3}{\mu_1\beta_3}w^2=\frac{\gamma_3}{\beta_3},
$$
where
$$
\alpha_1=\mu_1+\mu_3+\mu_2\frac{{\rho}^2_{2}}{{\rho}^2_{1}}, \ \
\beta_1=\mu_1+(\mu_2+\mu_3)\frac{{\rho}^2_{2}}{{\rho}^2_{1}}, \ \
\gamma_1=\mu_1(\mu_1+\mu_3)-\mu_2(\mu_2+
\mu_3)\frac{{\rho}^2_{2}}{{\rho}^2_{1}};
$$
$$
\alpha_2=\mu_1+\mu_2+\mu_3\frac{{\rho}^2_{3}}{{\rho}^2_{1}}, \
\beta_2=\mu_1(\mu_1+\mu_2)-\mu_3(\mu_2+\mu_3)\frac{{\rho}^2_{3}}{{\rho}^2_{1}},
\ \gamma_2=\mu_1+(\mu_2+\mu_3)\frac{{\rho}^2_{3}}{{\rho}^2_{1}};
$$
$$
\alpha_3=-\mu_3(\mu_1+\mu_3)\frac{{\rho}^2_{3}}{{\rho}^2_{1}}
+\mu_2(\mu_1+\mu_2)\frac{{\rho}^2_{2}}{{\rho}^2_{1}}, \
\beta_3=\mu_3\frac{{\rho}^2_{3}}{{\rho}^2_{1}}
+(\mu_1+\mu_2)\frac{{\rho}^2_{2}}{{\rho}^2_{1}},
$$
$$
\gamma_3=(\mu_1+\mu_3)\frac{{\rho}^2_{3}}{{\rho}^2_{1}}
+\mu_2\frac{{\rho}^2_{2}}{{\rho}^2_{1}}.
$$

Under the conditions of Lemma 3,  the considerable  movement is
Lagrange unstable. Hence, in accordance with Lemma 2, variable
coefficients ${\rho}^2_{2}(\tau_k)/{\rho}^2_{1}(\tau_k)$ and
${\rho}^2_{3}(\tau_k)/{\rho}^2_{1}(\tau_k)$ satisfy equalities
(2.11) with $k\rightarrow\infty$.

Let us consider  the limit version  of equations (2.24) when
$\tau\in\{\tau_k\}$, $(k=1,2,3,\dots)$. Taking equalities (2.11)
into account,  in the limit case, on the base of (2.24) we obtain
equalities (2.18) -- (2.20). Since  the system (2.18) -- (2.20),
which is treated as a system of linear equations with respect to
variables $\left({{\rho}^2_{13}}/{{\rho}^2_{12}}\right)_{\infty}$
and $\left({{\rho}^2_{23}}/{{\rho}^2_{12}}\right)_{\infty}$, is
inconsistent, we conclude  that only one of equalities (2.18) --
(2.20) for considerable motion  is valid.

Lemma 3 is proved.  $\Box$

\section {A Theorem on Lagrange Stability}

Let us try to use the information obtained in the previous section
in order to carry out a qualitative analysis of the movement
equations. In this connection, it should stressed that distance
equations (1.4) from the first section contain the term
$$
\frac{{\rho}^2_{13}-{\rho}^2_{23}}{{\rho}^2_{12}}.
$$
Along with this fact, similar terms are contained in the left-hand
sides of equations (2.18)--(2.20), though, it is true in the limit
case where we assume that the movement under consideration is
Lagrange unstable. Hence, there is a point in considering a
hypothetical possibility of the  Lagrange unstable movement in the
case of obtained movement equations hoping that we obtain some
useful information about qualitative behavior of movements in the
system. To this end we represent movement equation (1.3) in the
form
$$
{\bm \rho}_{12}''=-(1-\mu_3)\frac{{\bm \rho}_{12}}{|{\bm
\rho}_{12}|^3}+\mu_3\left(-\frac{{\bm \rho}_{13}}{|{\bm
\rho}_{13}|^3}+ \frac{{\bm \rho}_{23}}{|{\bm
\rho}_{23}|^3}\right),
$$
$$
{\bm \rho}_{13}''= -(1-\mu_2)\frac{{\bm \rho}_{13}}{|{\bm
\rho}_{13}|^3} -\mu_2\left(\frac{{\bm \rho}_{12}}{|{\bm
\rho}_{12}|^3}+ \frac{{\bm \rho}_{23}}{|{\bm \rho}_{23}|^3}
\right), \eqno (3.1)
$$
$$
{\bm \rho}_{23}''= -(1-\mu_1)\frac{{\bm \rho}_{23}}{|{\bm
\rho}_{23}|^3}+\mu_1\left(\frac{{\bm \rho}_{12}}{|{\bm
\rho}_{12}|^3}- \frac{{\bm \rho}_{13}}{|{\bm
\rho}_{13}|^3}\right).
$$
Equations (3.1) are more appropriate for our further purposes,
though equations (1.3) will be still considered as basic ones.

{\bf Theorem 1}. {\it Let ${\bm \rho}(\tau) =({\bm \rho}_{1},{\bm
\rho}_{2}, {\bm \rho}_{3})^T$ be a distal  movement of system
(1.3) that belongs to the set
$$
\Omega=\left\{({\bm \rho},{\bm \rho}'): T-U=h<0\right\}.
$$
Then, if masses $\mu_i (i=1,2,3)$ are different and one of the
pairs of the mass points is Hill stable, then the movement under
study is Lagrange stable}.

{\bf Proof.} Without loss of generality we can assume that
the pair $(\mu_1, \mu_2)$ is Hill stable.

Suppose that under the conditions of the theorem the movement
${\bm \rho}(\tau) =({\bm \rho}_{1},{\bm \rho}_{2}, {\bm
\rho}_{3})^T$ is  Lagrange unstable. Then there exist a sequence
$\{\tau_k\}$ $(k=1,2,3,\dots)$ such that
$$
\lim_{k \to \infty} \tau_k = \infty, \ \ \ \lim_{k \to \infty} \
\sum^3_{i<j}{\rho}^2_{ij} (\tau_k)= \infty. \eqno(3.2)
$$

Let us consider the function
$$
V={\bm \rho}_{12}{\bm \rho}_{13}'-\frac{1}{2}({\bm \rho}_{12}{\bm
\rho}_{13})', \eqno(3.3)
$$
which is formed on the base of the structure of the system of
equations (1.4). Its derivative with respect to the vector field,
which is determined by equations (3.1), has the form
$$
V'=({\bm \rho}_{12}{\bm \rho}_{13}')'-\frac{1}{2}({\bm
\rho}_{12}{\bm \rho}_{13})''=
\frac{1}{2}\left\{-\frac{\mu_2}{|{\bm
\rho}_{12}|}+(1-\mu_3)\frac{{\bm \rho}_{12}{\bm \rho}_{13}}{|{\bm
\rho}_{12}|^3}-\right.
$$
$$
\left.- \left[\frac{{\bm \rho}_{12}{\bm \rho}_{13}}{|{\bm
\rho}_{13}|^3}+(\mu_2{\bm \rho}_{12}+\mu_3{\bm
\rho}_{13})\left(-\frac{{\bm \rho}_{13}}{|{\bm \rho}_{13}|^3}+
\frac {{\bm \rho}_{23}}
{|{\bm\rho}_{23}|^3}\right)\right]\right\}. \eqno(3.4)
$$
Noticing that
$$
-\frac{\mu_2}{|{\bm \rho}_{12}|}+(1-\mu_3)\frac{{\bm
\rho}_{12}{\bm \rho}_{13}}{|{\bm \rho}_{12}|^3}=\frac{1}{2|{\bm
\rho}_{12}|}\left[(\mu_1-\mu_2) +
(\mu_1+\mu_2)\frac{{\rho}^2_{13}-{\rho}^2_{23}}{{\rho}^2_{12}}\right],
$$
we can rewrite equality (3.4) in the form
$$
V'=\frac{1}{2}\left\{\frac{1}{2|{\bm
\rho}_{12}|}\left[(\mu_1-\mu_2) +
(\mu_1+\mu_2)\frac{{\rho}^2_{13}-{\rho}^2_{23}}{{\rho}^2_{12}}\right]-
\right.
$$
$$
\left.- \left[\frac{{\bm \rho}_{12}{\bm \rho}_{13}}{|{\bm
\rho}_{13}|^3}+(\mu_2{\bm \rho}_{12}+\mu_3{\bm
\rho}_{13})\left(-\frac{{\bm \rho}_{13}}{|{\bm \rho}_{13}|^3}+
\frac {{\bm \rho}_{23}}
{|{\bm\rho}_{23}|^3}\right)\right]\right\}. \eqno(3.5)
$$
Assuming that the movement under study is  Lagrange unstable and
taking into account equalities (3.2) and also equalities
(2.18)--(2.20), on the base of (3.5) we obtain in the limit case
that
$$
(V')_{\infty}^{(1)}=\frac{1}{4|{\bm
\rho}_{12}|}\frac{\mu_1-\mu_2}{\mu_3}, \eqno(3.6)
$$
$$
(V')_{\infty}^{(2)}=-\frac{1}{4|{\bm
\rho}_{12}|}\frac{\mu_2}{\mu_3}, \eqno(3.7)
$$
$$
(V')_{\infty}^{(3)}=\frac{1}{4|{\bm
\rho}_{12}|}\frac{\mu_1}{\mu_3}. \eqno(3.8)
$$
The upper indices $1, 2, 3$ in the left-hand sides of equalities
(3.6)--(3.8) mean that instead of
$$
\frac{{\rho}^2_{13}-{\rho}^2_{23}}{{\rho}^2_{12}}
$$
in the right-hand side of equality (3.5) we substitute expressions
that are determined by right-hand sides of equalities (2.18),
(2.19), (2.20) respectively.

First let us consider equality (3.6), for which we assume that
$\mu_1>\mu_2$ and hence, we assume that the right-hand side of
equality (3.6) is positive. As a consequence of this fact, on the
base of continuity of the right-hand side of equality (3.5) we can
conclude that, for the sequence $\{\tau_k\}$, there is a
sufficiently large number $s$  such that the inequality
$$
V'\mid_{\tau\in \{\tau_k\}}\geq \delta_1 \ \  \forall k\geq s, \ \
0<\delta_1=\mbox{const}, \  \delta_1 < \frac{1}{4|{\bm
\rho}_{12}|}\frac{\mu_1-\mu_2}{\mu_3} \eqno(3.9)
$$
takes place for $k\geq s$.
In accordance with conditions of the theorem, the movement under study is distal, and hence velocities of mass points are bounded.
From this fact we can conclude that there is a sequence of time intervals with growing lengths
$$
\{T_j\}=[\tau_{s+j}-\tau_{n_{j}}], \  \tau_{s+j}\in \{\tau_k\}, \
j=1,2,3,\dots,\ \ \tau_{n_{j}}<\tau_{s+j}, \ \
n_{1}<n_{2}<n_{3}\dots,
$$
for which we have the inequality
$$
V'\geq \delta_1 \ \  \forall \tau \in \{T_j\}. \eqno(3.10)
$$

By integrating (3.10), we obtain the inequality
$$
V\mid_{\tau_1}^{\tau} \geq \delta_1(\tau-\tau_1), \ \ \tau >
\tau_1, \ \ [\tau_1, \tau]\subseteq \{T_j\}, \eqno(3.11)
$$
which can be further rewritten in the form
$$
-\frac{1}{2}({\bm \rho}_{12}{\bm \rho}_{13})'\mid^{\tau} \geq
-\frac{1}{2}({\bm \rho}_{12}{\bm \rho}_{13})'\mid_{\tau_1} - {\bm
\rho}_{12}{\bm \rho}_{13}'\mid_{\tau_1}^{\tau} +
\delta_1(\tau-\tau_1). \eqno(3.12)
$$
The product  $-{\bm\rho}_{12}{\bm\rho}_{13}'\mid_{\tau_1}^{\tau}$
is bounded on $R_\tau$ due to conditions of the theorem.
Therefore, by replacing it with a certain relevant constant
$\delta_2 > 0$, we can strengthen equality (3.12):
$$
-\frac{1}{2}({\bm \rho}_{12}{\bm \rho}_{13})'\mid^{\tau} >
-\frac{1}{2}({\bm \rho}_{12}{\bm \rho}_{13})'\mid_{\tau_1} -
\delta_2 + \delta_1(\tau-\tau_1). \eqno(3.13)
$$

By integrating inequality (3.13), we obtain
$$
-\frac{1}{2}({\bm \rho}_{12}{\bm \rho}_{13})\mid_{\tau_1}^{\tau}>
-\frac{1}{2}({\bm \rho}_{12}{\bm
\rho}_{13})'\mid_{\tau_1}(\tau-\tau_1)- \delta_2(\tau-\tau_1)+
\frac{\delta_1}{2}(\tau-\tau_1)^2. \eqno(3.14)
$$

Let us set $\tau_1=\tau_{n_{j}}$, $\tau=\tau_{s+j}$ in inequality
(3.14) and rewrite it in the form
$$
-\frac{1}{2}({\bm \rho}_{12}{\bm \rho}_{13})\mid_{\tau=\tau_{s+j}}
+ \frac{1}{2}({\bm \rho}_{12}{\bm
\rho}_{13})\mid_{\tau_{1}=\tau_{n_{j}}}
>
$$
$$
> (\tau_{s+j}-\tau_{n_{j}}) \left\{-\frac{1}{2}({\bm \rho}_{12}{\bm
\rho}_{13})'\mid_{\tau_{1}=\tau_{n_{j}}}- \delta_2+
\frac{\delta_1}{2}(\tau_{s+j}-\tau_{n_{j}})\right\}. \eqno(3.15)
$$
The terms
$$
\frac{1}{2}({\bm \rho}_{12}{\bm
\rho}_{13})\mid_{\tau_{1}=\tau_{n_{j}}}, \ \ -\frac{1}{2}({\bm
\rho}_{12}{\bm \rho}_{13})'\mid_{\tau_{1}=\tau_{n_{j}}}
$$
in (3.15) correspond to finite time points $\tau_{1}=\tau_{n_{j}}$
such that the sum $\sum^3_{i<j}{\rho}^2_{ij}(\tau_{n_{j}})$ reach
a critical value at which we have
$$
V'\mid_{\tau_{1}=\tau_{n_{j}}}\geq \delta_1.
$$
Hence, the quantities
$$
\frac{1}{2}({\bm \rho}_{12}{\bm
\rho}_{13})\mid_{\tau_{1}=\tau_{n_{j}}}, \ \ -\frac{1}{2}({\bm
\rho}_{12}{\bm \rho}_{13})'\mid_{\tau_{1}=\tau_{n_{j}}}
$$
in inequality (3.15) can be always chosen in such a way that they
are finite. Relating to this fact, it is appropriate for us to
rewrite inequality (3.15) in the form
$$
-\frac{1}{2}({\bm \rho}_{12}{\bm \rho}_{13})\mid_{\tau=\tau_{s+j}}
> -  \frac{1}{2}({\bm \rho}_{12}{\bm
\rho}_{13})\mid_{\tau_{1}=\tau_{n_{j}}} +
$$
$$
+ (\tau_{s+j}-\tau_{n_{j}})\left\{-\frac{1}{2}({\bm \rho}_{12}{\bm
\rho}_{13})'\mid_{\tau_{1}=\tau_{n_{j}}}- \delta_2+
\frac{\delta_1}{2}(\tau_{s+j}-\tau_{n_{j}})\right\}. \eqno(3.16)
$$
In accordance with (3.2) and the definition of time points
$\tau_{n_{j}}$, the length of the interval
$[\tau_{s+j}-\tau_{n_{j}}]$ tends to infinity as $j \to \infty$.
Hence, the right-hand side of inequality (3.16) tends to infinity
as well.

Now let us analyze the left-hand side of inequality (3.16) in a
more detailed way. To this end we note that
$$
{\bm \rho}_{12}{\bm \rho}_{13}=\frac{1}{2}({\bm \rho}_{12}^2+ {\bm
\rho}_{13}^2-{\bm \rho}_{23}^2),
$$
and represent it in the form
$$
-\frac{1}{2}({\bm \rho}_{12}{\bm
\rho}_{13})\mid_{\tau=\tau_{s+j}}= -\frac{1}{4}\left[{\bm
\rho}_{12}^2+ {\bm \rho}_{12}^2\frac{({\bm \rho}_{13}^2-{\bm
\rho}_{23}^2)}{{\bm \rho}_{12}^2}\right]_{\tau=\tau_{s+j}}.
\eqno(3.17)
$$
As $j$ tends to infinity, by equality (3.13) the terms inside the square brackets tend to the expression
$$
\left[{\bm \rho}_{12}^2+\frac{(\mu_1 - \mu_2)}{\mu_3}{\bm
\rho}_{12}^2\right]. \eqno(3.18)
$$
Thus, in accordance with our assumption $\mu_1 > \mu_2$, the
left-hand side of inequality (3.16) tends to a negative value as
$j \to \infty$. We arrive to a contradiction.

So, if the image point belongs to hyperbola (3.13) and $\mu_1 > \mu_2$, then the assumption on outgoing behavior of the movement ${\bm \rho}(\tau) =({\bm \rho}_{1},{\bm \rho}_{2}, {\bm \rho}_{3})^T$ is not true.

In an absolutely similar way we can obtain a contradiction in the
case where equality (3.8) is satisfied. Note only the fact that an
analogue of expression (3.18) in this case is the expression
$$
\left[{\bm \rho}_{12}^2+
\frac{\mu_2\mu_3^2+(\mu_1+\mu_3)(\mu_1+\mu_2)^2}
{(\mu_1+\mu_2)\mu_3}{\bm \rho}_{12}^2\right].
$$

Now consider equation (3.6) in the case where $\mu_1 < \mu_2$, and
hence, its right-hand side is negative. In this case, similarly to
the case that was studied above, due to continuity of the
right-hand side of equality (3.5) we can assert for the sequence
$\{\tau_k\}$ that there exist a sufficiently large number $s^*$
such that the inequality
$$
V'\mid_{\tau\in \{\tau_k\}}\leq - \delta_1^* \ \  \forall \  k\geq
s^*, \ \ 0<\delta_1^*=\mbox{const},  \  \delta_1^* <
\frac{1}{4|{\bm \rho}_{12}|}\frac{|\mu_1-\mu_2|}{\mu_3}
\eqno(3.19)
$$
takes place for $k\geq s^*$.
From this, by distality of the motion, we can conclude similarly to the case studied above that there exist a sequence of time intervals
$$
\{T^*_j\}=[\tau_{s^*+j}-\tau^*_{n_{j}}],  \tau_{s^*+j}\in
\{\tau_k\},  j=1,2,3,\dots, \tau^*_{n_{j}}<\tau_{s^*+j},
n_{1}<n_{2}<n_{3}\dots,
$$
with growing lengths for which the inequality
$$
V'\leq - \delta_1^* \ \  \forall \  \tau \in \{T^*_j\} \eqno(3.20)
$$
is satisfied.

By using almost literally the same scheme of arguments that was
used for equality (3.6) in the case where $\mu_1 > \mu_2$, we
arrive to an analogue of inequality (3.16):
$$
-\frac{1}{2}({\bm \rho}_{12}{\bm
\rho}_{13})\mid_{\tau=\tau_{s^*+j}} < -  \frac{1}{2}({\bm
\rho}_{12}{\bm \rho}_{13})\mid_{\tau_{1}=\tau^*_{n_{j}}} +
$$
$$
+ (\tau_{s^*+j}-\tau^*_{n_{j}})\left\{-\frac{1}{2}({\bm
\rho}_{12}{\bm \rho}_{13})'\mid_{\tau_{1}=\tau^*_{n_{j}}} +
\delta^*_2 -
\frac{\delta^*_1}{2}(\tau_{s^*+j}-\tau^*_{n_{j}})\right\},
0<\delta^*_2=\mbox{const}. \eqno(3.21)
$$
Due to (3.18), we can conclude that, as $j\to \infty$, the
left-hand side of inequality (3.21) tends to a bounded value and
the right-hand side tends to minus infinity. Hence, we arrive to a
contradiction.

Thus, the assumption on the  Lagrange instability  of the movement
under study is also not true in the case where the image point
belongs to hyperbola (3.13) as  $\mu_1 < \mu_2$.

Finally, it remains to consider the case where equality (3.7) is
satisfied. In this case, we can apply the arguments that were used
for equation (3.6) under the condition $\mu_1 < \mu_2$. It should
be note only the fact that an analogue of expression (3.18) in
this case will be represented by the expression
$$
\left[{\bm \rho}_{12}^2 - \frac{\mu_1\mu_3^2+(\mu_2+
\mu_3)(\mu_1+\mu_2)^2} {(\mu_1+\mu_2)\mu_3}{\bm \rho}_{12}^2\right].
$$

Thus, if we assume that the movement under study is Lagrange
unstable, then we arrive to a contradiction in all three cases
where equalities (3.6)--(3.8) take place. This contradiction give
us a possibility to conclude that the theorem is true. $\Box$

{\bf Remark 1}. As it is implied by the structure of equations
(1.4) and the scheme of proof of Theorem 1, the Lagrange stability
remains to be true also in the case where only different masses
are ones that form a Hill stable pair. For the third particle, it
is admissible that its mass is equal to the mass of a particle
from the Hill stable pair.

{\bf Remark 2}. If we take into account the fact that
$$
V={\bm \rho}_{12}{\bm \rho}_{13}'-\frac{1}{2}({\bm \rho}_{12}{\bm
\rho}_{13})'=
$$
$$
= ({\bm \rho}_{13} - {\bm \rho}_{23}){\bm \rho}_{13}'-
\frac{1}{4}({\bm \rho}_{12}^2+ {\bm \rho}_{13}^2-{\bm
\rho}_{23}^2)' = - {\bm \rho}_{23}{\bm \rho}_{13}'-
\frac{1}{4}({\bm \rho}_{12}^2 - {\bm \rho}_{13}^2-{\bm
\rho}_{23}^2)',
$$
then we can consider the derivative of the function
$$
V^*=- {\bm \rho}_{23}{\bm \rho}_{13}'- \frac{1}{4}({\bm \rho}_{12}^2 - {\bm \rho}_{13}^2-{\bm \rho}_{23}^2)'
$$
with respect to the vector field that is determined by equations
(1.4). However the function $V$ in the form (3.3) is more
appropriate. It is the function $V$ in the form (3.3) which is
predetermining the use of equations (3.1), though in the
construction of the function $V$ we are based on the structure of
the system of equations (1.4).

\section {On hyperbolic-elliptic and parabolic-elliptic final evolutions}

As it is known \cite {Chazy}, hyperbolic-elliptic and
parabolic-elliptic final evolutions are accompanied by a motion of
a bounded pair of particles and the third outgoing remote
particle. In this case, we can apply Lemma 2 in order to conclude
that relations (2.11) take place.

By using the Jacobi decomposition, we can represent the motion of
the bounded pair in the following convenient form:
$$
{\bf r}''=-(1-\mu_3)\frac{{\bf r}}{|{\bf r}|^3}+\mu_{3} \left(
\frac{{\bf R}-{\mu_{1}{\bf r}}/(\mu_{1}+\mu_{2})}{|{\bf
R}-{\mu_{1}{\bf r}}/(\mu_{1}+\mu_{2})|^3}- \frac{{\bf
R}+{\mu_{2}{\bf r}}/(\mu_{1}+\mu_{2})}{|{\bf R}+{\mu_{2}{\bf
r}}/(\mu_{1}+\mu_{2})|^3} \right). \eqno(4.1)
$$
Here, as it is usual, we have ${\bf r}={\bm \rho}_{12}$ and  $|\bf
R|$ denotes the distance from the third mass point to the center
of masses of the pair $(\mu_1,\mu_2)$. As we can see, vector
equation (4.1) represents the two-body problem with a decreasing
perturbation since the third particle is outgoing.

Since $|{\bf R}|\rightarrow\infty$, we see that ${\bf r}(\tau)$
tends to the elliptic Kepler motion with the relevant limit
integrals of the motion \cite{March}:
$$
\frac{\mu_{1}\mu_{2}}{\mu_{1}+\mu_{2}}\frac{{\bf
v}^2}{2}-\frac{\mu_{1}\mu_{2}}{|{\bf r}|}=h_{r}(\tau)\rightarrow
h_{r\infty}<0;    \eqno(4.2)
$$
$$
{\bf r}\times{\bf v}={\bf c}_{r}(\tau)\rightarrow{\bf
c}_{r\infty}.   \eqno(4.3)
$$

Let ${\bf r}_{a}(\tau)$ denote the asymptotic Kepler motion with
integrals $h_{r\infty}$ and ${\bf c}_{r\infty}$. In this case, in
accordance with \cite {March}, we have
$$
{\bf r}(\tau)-{\bf r}_{a}(\tau)= \left\{
\begin{array}{ll}
 O(\tau^{-2}), & {\bf c}_{r\infty}\neq {\bf 0}; \\
 O(\tau^{-4/3}), & {\bf c}_{r\infty}= {\bf 0}, \\
\end{array}
\right.           \eqno(4.4)
$$
if the evolution is hyperbolic-elliptic, and
$$
{\bf r}(\tau)-{\bf r}_{a}(\tau)=\left\{
\begin{array}{ll}
 O(\tau^{-1}), & {\bf c}_{r\infty}\neq {\bf 0}; \\
 O(\tau^{-2/3}), & {\bf c}_{r\infty}= {\bf 0} \\
\end{array}
\right.          \eqno(4.5)
$$
if the evolution is parabolic-elliptic.

It turns out that Theorem 1 provides a possibility to correct
equalities (4.4) and (4.5) respectively. In particular, we can
obtain the following statement.

{\bf Corollary of Theorem 1}. {\it Let masses $\mu_i$ $(i=1,2,3)$
in the three-body problem be different and $T-U=h<0$. Then in
cases of hyperbolic-elliptic and parabolic-elliptic final
evolutions, the following equalities are respectively valid:
$$
HE_{k}:\ {\bf r}(\tau)-{\bf r}_{a}(\tau)=O(\tau^{-4/3}), \ {\bf
c}_{r\infty}={\bf 0}, \eqno(4.6)
$$
$$
PE_{k}:\ {\bf r}(\tau)-{\bf r}_{a}(\tau)=O(\tau^{-2/3}), \ {\bf
c}_{r\infty}={\bf 0}, \eqno(4.7)
$$
i.e., going over to the limit, the modulus of the angular momentum
$|{\bf r}\times{\bf v}|$ of the bounded pair $(\mu_{1},\mu_{2})$
can not exceed a positive constant}.

{\bf Proof}. Let us suppose the contrary, ${\bf
c}_{r\infty}\neq{\bf 0}$, and consider the limit energy integral
for the pair $(\mu_{1},\mu_{2})$
$$
\frac{\mu_{1}\mu_{2}}{\mu_{1}+\mu_{2}}\frac{{\bf
v}^2}{2}-\frac{\mu_{1}\mu_{2}}{|{\bf r}|}=h_{r\infty}, \ \
h_{r\infty}<0,   \eqno(4.8)
$$
which, in its turn, can be rewritten in the form
$$
\frac{\mu_1\mu_2}{2}\left[\frac{1}{\mu_1+\mu_2}\left(|{\bf r}|'^2+
\frac{\left({\bf r}\times{\bf v}\right)^2}{|{\bf
r}|^2}\right)-\frac{2}{|{\bf r}|}\right]=h_{r\infty}. \eqno(4.9)
$$
Since $h_{r\infty}<0$, due to (4.9) we have
$$
\frac{1}{(\mu_1+\mu_2)}\frac{|{\bf c}_{r\infty}|^2}{|{\bf
r}|^2}-\frac{2}{|{\bf r}|}<0,
$$
and this implies
$$
|{\bf r}|>\frac{|{\bf c}_{r\infty}|^2}{2(\mu_1+\mu_2)}. \eqno(4.10)
$$

In accordance with inequality (4.10), we conclude that if ${\bf
c}_{r\infty}\neq{\bf 0}$, then hyperbolic-elliptic and
parabolic-elliptic final evolutions are accompanied by a distal
motion. However, according to Theorem 1, for $T-U=h<0$ the distal
motion with a fixed bounded pair is Lagrange stable. We obtain a
contradiction and this implies that the corollary is true. $\Box$

\section {Conclusion}

Summarizing the above represented results, we can state that the
key requirements of the proved theorem that provide Lagrange
stability are existence of a pair of points that are Hill stable
and distality of the movement. Unfortunately, the problem of
choice of initial conditions and parameters of the system that
provide the distal movements is still open. In this relation, it
is interesting to note that conditionally periodic motions, the
existence of which in the three-body problem is proved in the
Kolmogorov-Arnold-Moser theory, belong to the class of distal
motions. This means that Theorem 1 is constructive. Corollary of
Theorem 1 deepens our understanding of hyperbolic-elliptic and
parabolic-elliptic final evolutions in the three-body problem.

\newpage

\renewcommand \refname{References}

\end{document}